# Pulsed Laser Ejection of Single-Crystalline III-V Solar Cells From GaAs Substrates

Benjamin A. Reeves,[1] Myles A. Steiner,[2] Thomas E. Carver,[3] Ze Zhang,[4]
Aaron M. Lindenberg,[1,5] Bruce M. Clemens[1*]

1. Department of Materials Science and Engineering, Stanford University, Stanford, CA.

2. National Renewable Energy Laboratory, Golden, CO.

3. Stanford Nano Shared Facilities, Stanford University, Stanford, CA.

4. Department of Mechanical Engineering, Stanford University, Stanford, CA.

5. Stanford Institute for Materials and Energy Sciences, SLAC National Accelerator Laboratory, Menlo Park, CA 94305

* To whom correspondence should be addressed: bmc@stanford.edu

**Abstract**

Like many optoelectronics, the highest quality III-V solar cells start out as thin single-crystalline multilayers on GaAs substrates. Separating these device layers from their growth substrate enables higher performing devices and wafer reuse, both of which are critical for III-V solar cell viability in a terrestrial market. Here, we remove rigidly-bonded, lattice-matched, 16 mm$^2$ x 3.5 µm thick GaAs devices off a GaAs substrate using a 10 ns, unfocused Nd:YAG laser pulse. The pulse is selectively absorbed in a lower-bandgap, lattice-matched, crystalline layer below the device, driving a quasi-two dimensional ablation event that ejects the crystalline multilayer from the substrate. After minutes of selective wet-chemical etching and front contact deposition, our champion 0.1 cm$^2$ device showed a (17.4 ± 0.5) % power conversion efficiency and an open-circuit voltage of 1.07 V, using AM1.5 direct (1000 W m$^{-2}$) with no anti-reflection coating. We show that the performance is comparable to similar solar cells produced via conventional substrate dissolution processes. We discuss unique process characteristics and opportunities, such as the potential to separate wafer-sized thin film solar cells per laser pulse.

**Introduction**

Thin film transfer and wafer recovery processes are essential for manufacturing single crystal III-V solar cells. III-V substrates are typically two to three orders of magnitude thicker than the active photovoltaic layers,[1] and III-V wafer costs are high because, for example, III-V elements and compounds are not abundant.[2] They are also toxic, carcinogenic,[3] and fragile,[4] and III-V wafer manufacturing utilizes specialized, encapsulated crystal pulling techniques for relatively small wafer diameters and with high material losses.[4,5] Yet, all record-holding solar cells are made from single crystal III-V thin films that were grown on III-V substrates.[6-9] The proliferation of the highest-quality solar cells therefore depends, in part, on rapid and inexpensive processes that separate single crystal thin films from III-V substrates while preserving wafer surfaces for regrowth.[10,11]

Driven in part by demand for lower-cost III-V photovoltaics, various III-V single-crystalline thin film transfer solutions have been developed, all of which utilize some combination of epitaxy, mechanical peeling, and selective chemical etching. Wet-chemical etch-rate ratios between lattice-matched alloys can exceed $10^6$.[12,13] This etch selectivity is utilized to separate thin films from growth substrates, for example by laterally etching a 10 - 100 nm thick AlAs[14,15] or AlInP[16] layer grown between a GaAs substrate and the device layers. The lateral etches become diffusion-limited and stall unless the film is peeled away during etching, and then proceed only at order 1 - 10 mm hr$^{-1}$.[17] With spalling, a mechanical separation process, a tensile stressor layer on the thin film surface pulls open a lateral crack beneath the device layers.[18] Unfortunately, for (100) III-V substrates, cracks opening along a {100} plane will redirect onto the lower-energy {110} planes, and potentially into device layers and the substrate.[19] Variations on spalling and chemical liftoff processes seek to address crack-confinement,[20,21] stress management,[22] lateral etching rates,[17] and surface effects.[16]

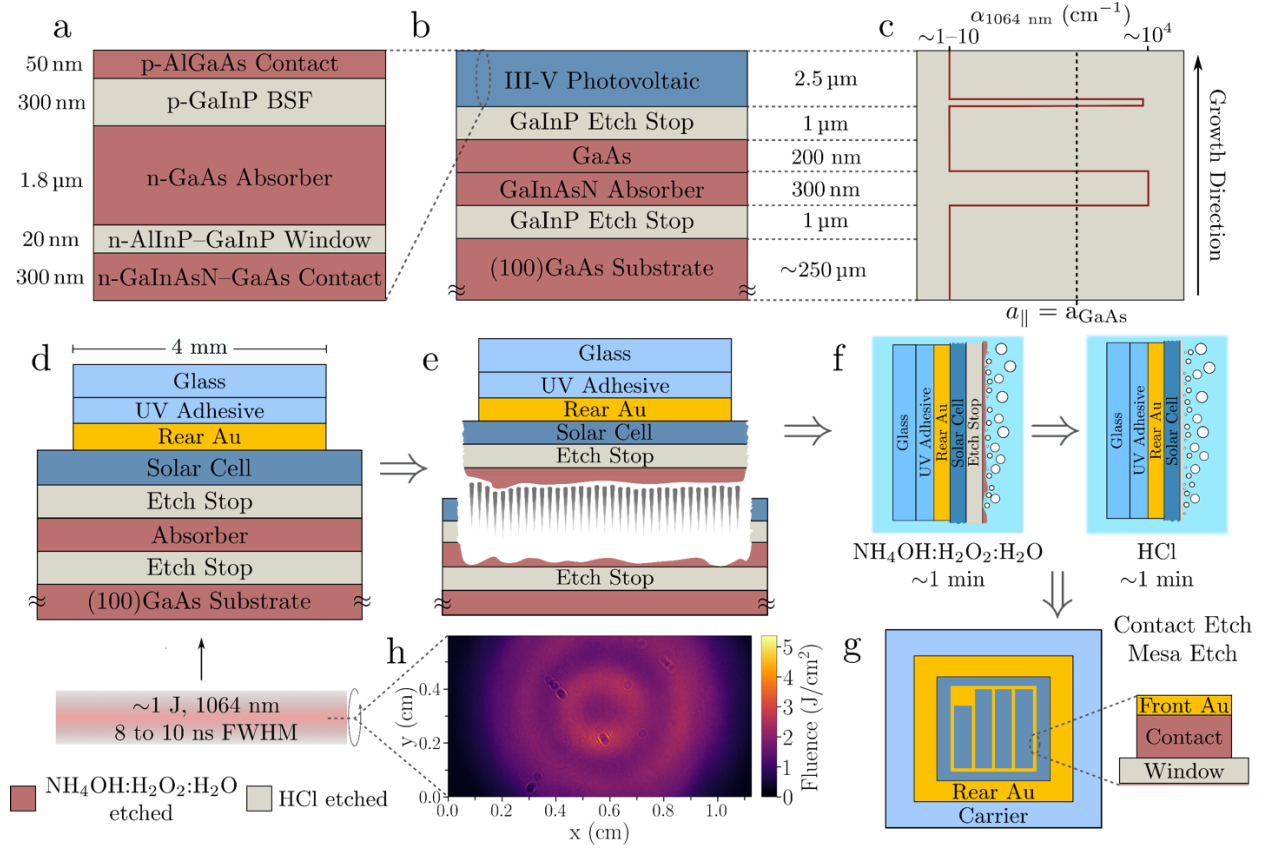

**Figure 1: Overview of device fabrication with pulsed-laser crystal ejection.** a) This work used GaAs-based, single-heterojunction, inverted, lattice-matched, III-V photovoltaics. b) The photovoltaic was synthesized in line with the layers used for device ejection and finishing (a lattice-matched, GaInP-GaInAsN-GaAs-GaInP multilayer). c) The relatively narrow, direct electronic bandgap of the GaInAsN absorber created a sharp absorption gradient for a Nd:YAG laser pulse inside of the crystal. d) After depositing a Au|Ti contact layer onto the p-AlGaAs, the Ti (not-shown) was bonded to glass, and the surrounding metal etched away. The specimen was then positioned into the path of a normally-incident Nd:YAG laser pulse. e) The laser pulse drove an ablation event inside of the crystal, which caused the photovoltaic device structure to eject from the substrate. f) After ejection, the melt debris and etch stop were etched away in minutes with room temperature, chemically selective etchants. Prior to etching, the glass was bonded to a larger glass carrier, to improve device handling (not shown for simplicity). g) After electroplating the Au grid, the exposed front GaInAsN contact layer was etched down to the AlInP window. The mesa was defined with photolithography and etched, similarly to f. h) Our laser spatial filter could produce single- or multi-mode laser pulse spatial profiles. Here, we used multi-mode pulses to maximize transmitted energy. The small Fresnel diffraction spots rotated around the profile when rotating imaging optics and are therefore not intrinsic to the laser spatial profile.

Here, we eject rigidly-bonded III-V photovoltaic devices from a GaAs wafer using a laser pulse (Fig. 1), by exploiting wavelength-dependent optical absorption $\alpha(\lambda)$ differences between conventional, lattice-matched III-V layers.[25,26] The device structure is a multilayer of III-V alloys, with sharp, optoelectronic property gradients, and active layers are single crystals of order 1 nm to 1 μm thick. Individual layers provide critical functionalities,[27-29] such as surface passivation with indirect, wide-bandgap III-V windows, carrier-selective transport, Ohmic contacting, and now, additionally, a degree of freedom to optically split the active layers from the GaAs substrate before minutes-long wet etching and device finishing. We compare our prototype, laser-ejected solar cells to specimens produced at the National Renewable Energy Laboratory (NREL) from mature fabrication processes and comparable metalorganic vapor phase epitaxy (MOVPE) growth parameters.

**Results and Discussion**

We synthesized standard III-V thin films via MOVPE at NREL (Fig. 1a, 1b, Methods, S1). Beginning with a single-side polished (100) GaAs substrate, we grew a 1 μm thick $Ga_{0.49}In_{0.51}P$ etch stop, 300 nm $Ga_{.93}In_{.07}As_{.98}N_{.02}$ absorber layer, 200 nm GaAs spacing layer, and another 1 μm thick $Ga_{0.49}In_{0.51}P$ etch stop (all thicknesses and compositions are nominal). Growth continued with an inverted rear-heterojunction III-V photovoltaic cell,[30] consisting of an n-$Ga_{.97}In_{.03}As_{.99}N_{.01}$ front contact layer, n-$Al_{0.52}In_{0.48}P$|$Ga_{0.49}In_{0.51}P$ window layers, n-GaAs absorber, p-$Ga_{0.49}In_{0.51}P$ back surface field (BSF), and p-$Al_{0.3}Ga_{0.7}As$ rear contact layer. Relative to mature baseline devices, the GaInP-GaInAsN-GaAs-GaInP layer set was the only growth modification necessary for laser ejection. The two GaInAsN layers were slightly different: The laser absorber layer had ~2% N on the group-V sub-lattice and ~7% In on the group-III sub-lattice to lattice-match the alloy, while the front contact layer had ~1%N and 3% In to lattice-match. Photoluminescence and X-ray diffraction results from the $Ga_{0.93}In_{0.07}As_{0.98}N_{0.02}$ absorber layer were consistent with a pseudomorphic alloy, with a bandgap of approximately 1.07 eV (S2), which is 0.1 eV less than the laser's 1.17 eV photon energy. As shown in Fig. 1c, based on typical behaviors of direct-band gap III-V alloys,[31] the GaInAsN absorber (and GaInAsN front contact) likely had absorption coefficients $\alpha$(1064 nm) of about $10^4$ cm$^{-1}$. However, $\alpha$(1064 nm) for the other compounds were orders of magnitude lower because their electronic band gaps were greater than the photon energy, although their absorption was not zero due to the non-linear processes that occur at our order 100 MW cm$^{-2}$ intensities.[32,33] For all layers, the lattice-matching preserved the in-plane lattice parameter $a_\parallel = a_{GaAs}$, preventing strain-induced defects such as misfit dislocations.

After MOVPE, we prepared the specimen for crystal ejection. The substrate was single-side polished for better in-situ monitoring of the curvature and stress during MOVPE, but because the laser pulse would travel through the substrate, the substrate was polished after epitaxy to produce a specular surface and an approximate thickness of 250 μm. The specular surface was protected with photoresist until laser processing. The AlGaAs was stripped of surface oxides via wet etching, followed by e-beam evaporation of a 500 nm rear Au reflective contact, then a 40 nm Ti adhesion layer. 4 mm x 4 mm glass coverslips were bonded to the Ti with UV-curing adhesive and direct-write UV photolithography.

After removing the surrounding Ti and Au with wet etching, the contacted and mounted device layers were inserted normal the optical path of a commercial, Q-switched, Nd:YAG laser, with the polished wafer surface facing the laser. The laser produced nominally 1064 nm wavelength, 8 - 10 ns full-width-at-half-maximum (FWHM) pulses (Figs. 1d and 1h, S3). Sharp variations in the laser pulse spatial intensity profile were minimized with a vacuum-pinhole spatial filter, yielding pulses with the characteristic spatial fluence profile shown in Fig. 1h, and pulse energies about $1.1 \pm 0.1$ J (the small diffraction spots in Fig. 1f rotated with imaging optics, and are therefore imaging artifacts). After course alignment by transmission imaging of highly attenuated pulses, we removed attenuation filters and ejected the specimen. We observed single-shot specimen ejection (S4), but if a single pulse did not eject a specimen—due to misalignment or pulse energy fluctuation, for example—the specimen was translated by order 100 μm before firing another pulse.

The ejected specimen's glass was adhered to a glass carrier for selective wet etching and device finishing (Figs. 1f-1g). The melt debris and GaInP etch stop were removed with their respective selective etchant (the substrate could be treated analogously, S4). We then created the front metal contact grid by direct-write photolithography, and electroplating of a thin Ni adhesion layer and nominally 2 μm of Au. The exposed GaInAsN contact layer was removed via wet etching, and finally, a mesa was defined via photolithography and additional selective etching.

The devices' photovoltaic performance was measured at NREL. Fig. 2 shows the testing configuration, current density vs. voltage (J-V) curves under the AM1.5 direct spectrum at 1000 W/m$^2$, dark J-V, and external quantum efficiency data for our best cells. We compare our laser-ejected devices to similar baseline cells grown in the same MOVPE reactor, and fabricated via NREL's mature GaAs processes, where the substrate was chemically etched away.[34,35] As summarized in Table 1, our prototype ejected devices demonstrated excellent performance metrics, near parity with the baseline. With no anti-reflection coatings, our highest performing cell showed a conversion efficiency η=(17.4 ± 0.5)%, short circuit current density $J_{sc}$=(19.8 ± 0.6) mA/cm$^2$, open circuit voltage $V_{oc}$=1.07 V, fill factor FF=0.82, and device area A=0.105 cm$^2$. The external quantum efficiency data also show comparable performance, with characteristic, sharp electronic band edges, and Fabry-Perot interference fringes from the single-crystalline uniform multilayers and highly reflective back mirror. The quantum efficiency differences between the etched and ejected samples are on the order of a few percent at all wavelengths. The dark J-V shows a difference between the etched and ejected samples. Both sets of samples exhibit normal n=1 diode behavior at high voltage and n=2 behavior at low voltage, but the ejected samples show shunt-like behavior at intermediate voltages, that likely explains the reduced fill factors. The origin of the shunt is unclear at this time, but it would appear to be a non-linear shunt that does not affect the slope of the J-V curve (Fig. 2b) near 0 V, as in a more typical shunt.

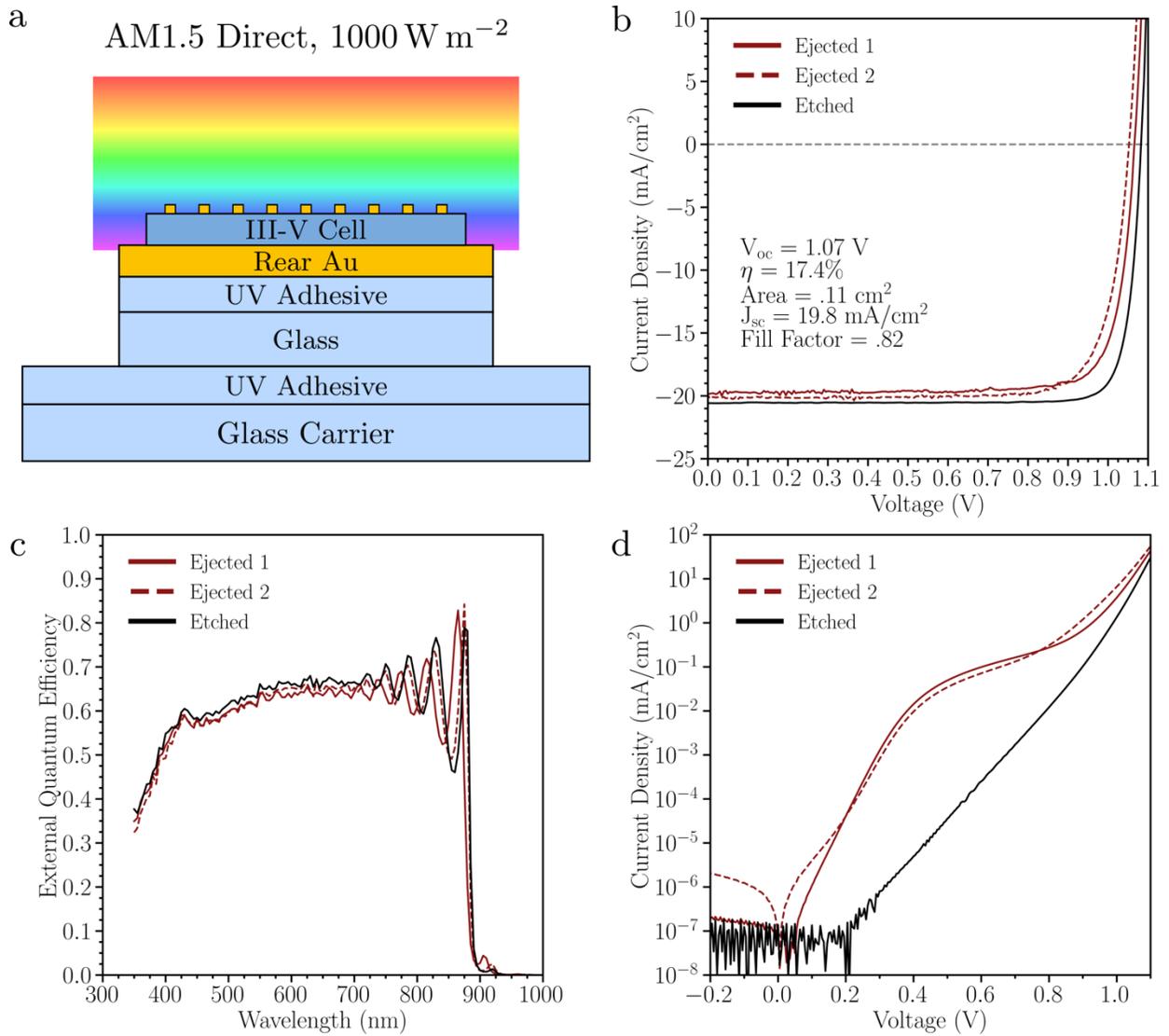

**Figure 2: Results from optoelectronic device characterization.** a) J-V tests on the final ejected device structures were measured under AM1.5 direct, 1000 W/m² conditions at 25°C. b) The ejected and etched cells showed excellent, comparable performance. The performance metrics for our best-performing ejected cell are noted in the figure and in the table below. c) External quantum efficiency was comparable between etched and ejected cells. d) Dark curves for the ejected cells showed a non-linear shunt for the ejected devices.

**Table 1: A summary of etched and ejected device performance metrics.** Photovoltaic performance metrics for the two laser-ejected cells, and the baseline cell whose substrate was etched away with $NH_4OH:H_2O_2$.

| Device | η (%) | $J_{sc}$ (mA/cm$^2$) | $V_{oc}$ (V) | FF | Area (cm$^2$) |
|---|---|---|---|---|---|
| Ejected 1 | 17.4 | 19.8 | 1.07 | 0.821 | 0.105 (measured) |
| Ejected 2 | 17.0 | 19.9 | 1.05 | 0.810 | 0.101 (measured) |
| Etched | 19.2 | 20.6 | 1.08 | 0.863 | 0.1005 (nominal) |

We observed finished, ejected devices with bright field, Nomarski, and dark field microscopies. Despite our inhomogeneous, multi-mode laser pulses (Fig. 1h), we did not observe any thin film cracking or other bulk damage for the device dimensions that were readily ejectable with our laser system (Figs. 3a-c, S5). We did observe thin film damage near the ejected film edges and corners, especially after wet etching. During laser ejection, device corners were located within a few hundred microns of the sub-critical fluences near our pulse edges, and could mechanically tear during ejection. Furthermore, unlike the baseline cells, our ejected devices were elevated from the surface of the glass carrier (by the Ti-bonded glass). Hence, effects such as photoresist beading and adhesive swelling along the elevated edges complicated wet etching, photolithography, and electroplating. Larger sample sizes and statistical analyses were beyond the scope of the present study, but will be required to further improve performance, as well as understand any damage mechanisms, their effects, and appropriate mitigation strategies from within the vast process parameter space. At a minimum, future experiments should explore flat laser spatial profiles, with sufficient energy and diameter to extend beyond the edges of order 1 cm$^2$ to 100 cm$^2$ ejection specimens.

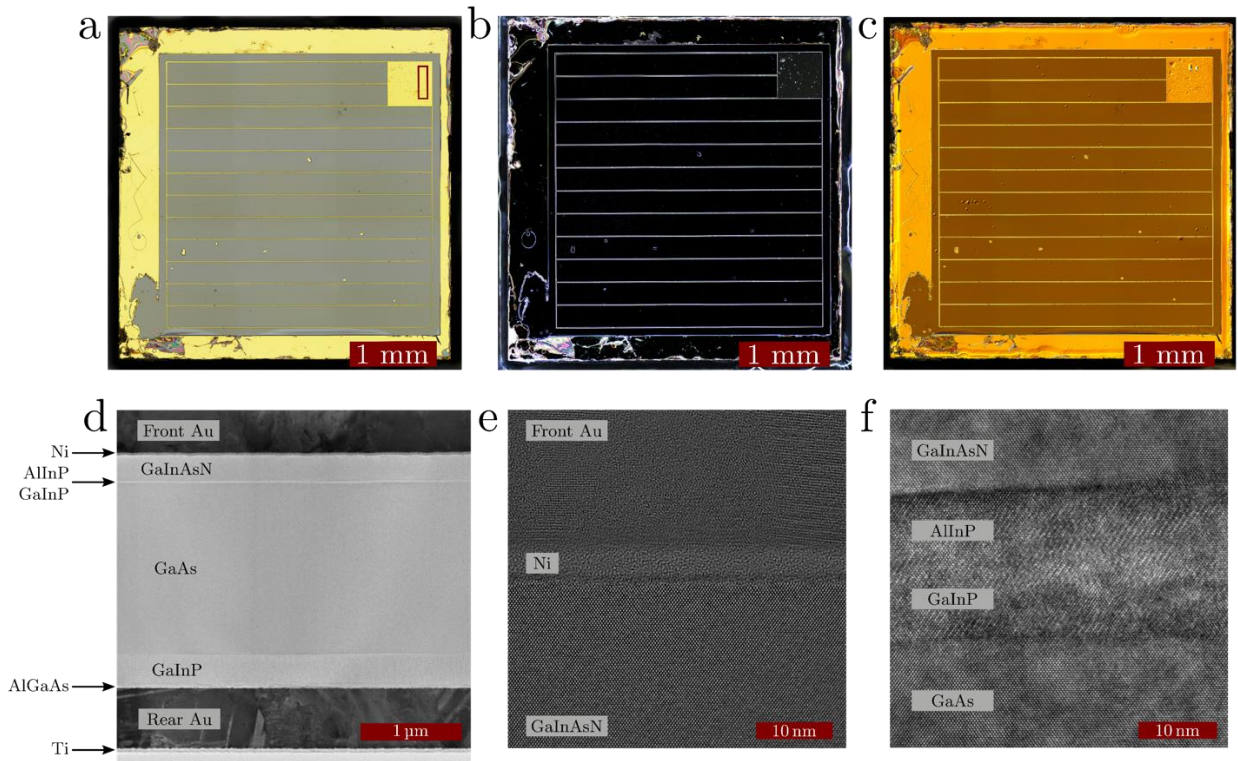

**Figure 3: Optical and TEM device characterization.** a) A bright-field optical micrograph of the finished cell. The rectangle over the front contact pad shows the characteristic ion-milling location for TEM specimens. Image stitching artifacts are seen in a-c. b) Dark field optical microscopy reveals a crack-free surface, as well as debris on the AlInP window. c) Differential interference contrast microscopy also shows a crack free surface, as well as electroplating roughness and rounded subsurface features not seen in a or b. d) A TEM cross section of an ejected cell, showing layer thicknesses and excellent material quality. The <100> direction is vertical and <110> is normal to the page. e, f) We found no evidence of misfit dislocations or melted interfaces while searching the lamella shown in d. The scale bars in e and f were calibrated by assuming GaAs atomic spacing in the micrographs.

Prior transmission electron microscopy (TEM) and X-ray diffraction work showed that GaAs thin films ejected from a GaAs substrate were free from misfit dislocations, and had the same structural quality.[23] We used TEM to search for dislocations and disordering in a laser-ejected, finished GaAs cell (Figs. 3d-f). Using an ejected solar cell from this batch that failed during photovoltaic testing, we removed a cross-sectional lamella from between the front Au contact pad and rear Au contact using a focused Ga ion beam and micromanipulators (Fig. 3a). The transmission electron micrographs revealed the single-crystalline multilayers characteristic of high-performing III-V photovoltaic cells. We found atomically-coherent interfaces and no misfit dislocations, a consequence of targeting III-V compositions that are lattice-matched to GaAs (and whose composition deviations were not sufficient to exceed critical stresses during growth).[35] These data also show that the crystalline thin films and precise interfaces persisted even though the layers were within about 1 μm of a fast, intracrystalline ablation transient. It is especially notable that both the rear Au contact and the front GaInAsN contact layer were intact after ejection, as these layers would absorb and reflect any energy transmitted during the first few ns of the laser pulse, before the absorber layer became opaque. These data support the conclusion that, for the bonding conditions, specimen design, laser system, and ejection scheme used in this work, the amplitude and duration of the thermomechanical transients during and after ejection do not necessarily damage the finished crystals.

Our solar cells were removed from their growth substrate with a ~10 ns laser pulse, and are the largest single-crystalline devices that we have ejected via pulsed laser energy. Damage was restricted to within approximately 1 μm of a pre-determined epitaxial interface, and the ejected film was only 3.5 μm thick; both of these dimensions are smaller than those we found reported for spalling.[20,21] We postulate that crystal ejection's processing space has at least seven important dimensions: (1) Laser wavelength, (2) pulse duration, (3) pulse energy, (4) laser spatial profile, (5) multilayer architecture (including metal reflectors, thicknesses, and the substrate's composition), (6) initial strain (intrinsic, as well as bonding-induced), and (7) surface treatments. Beyond successful ejection of these 4 mm x 4 mm glass|Ti|Au|III-V structures, we do not know the ultimate ejected-area limit at any point in the processing space, nor the minimum amount of material that must be sacrificed for ejection and device finishing. However, at least for 1 - 10 ns, 1064 nm, Nd:YAG lasers, wafer-scale pulses can be created at tens of Hz with commercial laser systems.[37] Coupled with the present results, and the accuracy of MOVPE, this suggests that it is already possible to eject lithographically-patterned arrays of mm-scale III-V devices, even if thermomechanical cracks or other issues emerge with larger, continuous thin films.

Various facets of this process demonstration motivate discussions about scalability and fundamental crystal creation limits. Our process consists of epitaxy, bonding, laser ejection via free space, and post-ejection finishing. In principle, our post-epitaxial surface never has to touch more than one substance, for example a gold contact. We ejected in air here, but could also use an inert atmosphere, or even a vacuum. The volume that forms between surfaces during ejection, at least at small displacements or in vacuum, will consist only of excited material from within the crystal. In other words, the ejection process only requires enough empty space for the crystals to separate, and then the new surfaces cool in their own, ultra-pure environment. To reveal a pristine interface, we chose to remove the debris and etch stops by rapid wet etching. These layers can be etched away *relatively* slowly via dilute mixtures, or at 1 to 10 µm per minute with standard, room temperature solutions. After removing the melt debris, another short etch reveals the final interface, e.g. the device's surface. But, for nearly all of this time, the etchant is not in contact with the device's front contact surface. No part of the front contact is exposed to the etchant until the final moments of the last etch, during the brief time that it takes to clear uneven etch-fronts and verify completion before rinsing. Coupled with the variety of flow, chemistry, temperature, and orientation options available to us by avoiding thick etches and chemical undercutting, it is reasonable to expect that this process can minimize reaction product surface contamination,[16] as well as surface roughness, especially for material-etchant pairs with small etch rate ratios. Hence, assuming that insoluble or otherwise persistent phases are avoidable, this process shows ideal characteristics for generalized III-V crystal splitting, bonding, and etching.

This process may also approach fundamental rate and volumetric processing limits for single-crystalline III-V optoelectronics. Our etch times were independent of device area, and we rapidly etched layers in vertical configurations. We did not peel thin films, use stressor layers, or need any of the concomitant equipment, material, or controls in our process environment. Therefore, we might minimize etch bath volumes, evaporation rates, bath contamination, cleanroom contamination, cleanroom volume, energy consumption, process interruption losses, and waste, while maximizing purity, recovery, recyclability, and production rate per volume. Our results suggest pathways to, for example, square-meters-per-second throughput from an order 10 $m^2$ commercial laser station and ~0.1 $m^3$ chemical baths, and competing across a variety of interesting fabrication metrics, such as peak watts solar per manufacturing volume per time.

It is remarkable that such a wide, thin, single-crystalline multilayer can be split with one laser pulse, via free space, without stressor layers, and with rigid (*or* flexible[23]) boundary conditions. Characterizing and understanding the intracrystalline transients is a complex, multidisciplinary problem, especially given the breadth of this processing space. Nevertheless, it is reasonable to guess that our crystals separated along the absorber plane on order 10 ns time scales. This would be orders of magnitude faster than speed-of-sound propagation times across even the modest length scales demonstrated here. Furthermore, we are essentially optically disassembling a (quasi) single crystal lattice everywhere at once across two dimensions. This implies that ultrafast electron-photon or electron-phonon coupling and local phase transformation times will set the ultimate speed limits for splitting crystals. Finally, because we applied selective optical absorption to split the crystal, we reserved our selective chemical etches for post-separation surface processing. We therefore demonstrated orthogonal processing operations to split crystals and expose interfaces. More generally, we can view wavelength-selective absorption as adding orthogonal dimensions to the traditional III-V processing space, and postulate higher-order pathways such as optically guided epitaxial spalling, void-assisted lateral etching, or simply the rapid chemical-mechanical bifurcation of a laser-ejected GaAs|AlAs|GaAs thin film. These processing considerations, and our facile demonstration of single-crystalline minority-carrier devices, imply that crystal ejection could play an important role in the development and proliferation of critical optoelectronics.

**Acknowledgements**

B.A.R. was supported by the Department of Defense and Air Force Office of Scientific Research through the National Defense Science & Engineering Graduate Fellowship Program. Part of this work was performed at the Stanford Nano Shared Facilities (SNSF), supported by the National Science Foundation under award ECCS-1542152. B.A.R. thanks Timothy Brand of Stanford's Ginzton Crystal Shop for crystal polishing, Michelle Young, Prof. Nick Rolston, and Stanford's Bent and Brongersma Groups for their generous support.  A.M.L. acknowledges support from the Department of Energy,  Office of Science, Basic Energy Sciences, Materials Sciences and Engineering Division, under Contract DE-AC02-76SF00515

This material is based upon work supported by the U.S. Department of Energy's Office of Energy Efficiency and Renewable Energy (EERE) under the Solar Energy Technologies Office Award Number 34358. The views expressed herein do not necessarily represent the views of the U.S. Department of Energy or the United States Government.

**Author Contributions**

Conceptualization, B.A.R. and B.M.C.; Investigation, B.A.R. (laser ejection, cell fabrication, XRD and PL), Z.Z. (TEM), M.A.S. (PV characterization, cell design, and epitaxy), and T.E.C. (evaporation); Writing – Original Draft, B.A.R.; Writing – Review & Editing, all authors; Funding Acquisition, B.A.R., M.A.S., A.M.L., and B.M.C. Resources, M.A.S. and A.M.L. Supervision: M.A.S., A.M.L., and B.M.C.**References**

1. Stringfellow, G. *Organometallic Vapor-Phase Epitaxy: Theory and Practice* (Academic Press Inc., 1989).

2. Abundance of Elements in the Earth's Crust and Sea. Edited by Rumble, J. *CRC Handbook of Chemistry and Physics* 102nd ed. (2021).

3. Zeng, C. *et al.* Ecotoxicity assessment of ionic As(III), As(V), In(III) and Ga(III) species potentially released from novel III-V semiconductor materials. *Ecotoxicology and Environmental Safety* **140**, 30–36 (2017).

4. Ghandhi, S. *VLSI Fabrication Principles: Silicon and Gallium Arsenide* 2ed (John Wiley & Sons, 1994).

5. Rudolph, P & Jurisch, M. Bulk growth of GaAs An overview. *Journal of Crystal Growth* **198-199**, 325–335 (1999).

6. Kayes, B. M. *et al.* 27.6% Conversion efficiency, a new record for single-junction solar cells under 1 sun illumination in *2011 37th IEEE Photovoltaic Specialists Conference (PVSC)*, 4–8.

7. Sasaki, K. *et al.* Development of InGaP/GaAs/InGaAs inverted triple junction concentrator solar cells. *AIP Conference Proceedings* **1556**, 22–25 (2013).

8. Chiu, P. T. *et al.* 35.8% space and 38.8% terrestrial 5J direct bonded cells in *2014 IEEE 40th Photovoltaic Specialist Conference (PVSC)*, 0011–0013.

9. Green, M. A. *et al.* Solar cell efficiency tables (Version 53). *Progress in Photovoltaics: Research and Applications* **27**, 3–12 (2019).

10. Horowitz, K. A., Remo, T. W., Smith, B. & Ptak, A. J. *A Techno-Economic Analysis and Cost Reduction Roadmap for III-V Solar Cells* NREL/TP-6A20-72103 (National Renewable Energy Lab. (NREL), Golden, CO (United States), 2018).

11. Ward, J. S. *et al.* Techno-economic analysis of three different substrate removal and reuse strategies for III-V solar cells. *Progress in Photovoltaics: Research and Applications* **24**, 1284–1292 (2016).

12. Hjort, K. Sacrificial etching of III-V compounds for micromechanical devices. *Journal of Micromechanics and Microengineering* **6**, 370 (1996).

13. Yablonovitch, E., Gmitter, T., Harbison, J. P. & Bhat, R. Extreme selectivity in the liftoff of epitaxial GaAs films. *Applied Physics Letters* **51**, 2222–2224 (1987).

14. Konagai, M., Sugimoto, M. & Takahashi, K. High efficiency GaAs thin film solar cells by peeled film technology. *Journal of Crystal Growth* **45**, 277-280 (1978).
13

# Materials and Methods: Pulsed Laser Ejection of Single-Crystalline III-V Solar Cells From GaAs Substrates

Benjamin A. Reeves,[1] Myles A. Steiner,[2] Thomas E. Carver,[3] Ze Zhang,[4] Aaron M. Lindenberg,[1,5] Bruce M. Clemens[1*]

1. Department of Materials Science and Engineering, Stanford University, Stanford, CA.

2. National Renewable Energy Laboratory, Golden, CO.

3. Stanford Nano Shared Facilities, Stanford University, Stanford, CA.

4. Department of Mechanical Engineering, Stanford University, Stanford, CA.

5. Stanford Institute for Materials and Energy Sciences, SLAC National Accelerator Laboratory, Menlo Park, CA 94305

* To whom correspondence should be addressed: bmc@stanford.edu

## Thin Film Epitaxy

Thin films were synthesized in NREL's custom-built, vertical flow, atmospheric pressure, metalorganic vapor phase epitaxy (MOVPE) system. Group III precursors were trimethylgallium, triethylgallium, trimethylaluminum, and trimethylindium, and group V precursors were arsine and phosphine. Dopant precursors were carbon tetrachloride (C), diethylzinc (Zn), dilute hydrogen selenide (Se), and disilane (Si). Nitrogen was sourced from dimethylhydrazine. Before each growth, the MOVPE chamber's quartz reactor tube and graphite susceptor were etched in aqua regia and rinsed in DI water. The chamber was then heated to 1000 °C, held for 2 min and cooled, in order to reduce O contamination.

Structures were grown on epi-ready, Si-doped (~3E18 cm$^{-3}$ based on manufacturer's specifications), (100)GaAs substrates with a 2° miscut towards the (111)B plane (i.e. As-terminated on the step-edges). The substrate was diamond-scribed and cleaved along {110} planes into a 2.5 cm x 2 cm growth piece, and then etched for two minutes in $NH_4OH : H_2O_2 : H_2O$ (2:1:10 by volume) and rinsed in DI water. This substrate piece was then loaded into the chamber under $N_2$. The carrier gas was switched to $H_2$, the reactor heated to 700 °C under $AsH_3$, and the substrate deoxidized for about 10 minutes.

The solar cell was grown inverted, and all reported thin film thicknesses and compositions are nominal. *In-situ* diagnostics were used to monitor the growth (Figure S1). We deposited a 100 nm GaAs seed layer and 1 µm of $Ga_{0.51}In_{0.49}P$ (lower stop-etch). The reactor was cooled to 570 °C to grow 50 nm of $Ga_{0.51}In_{0.49}P$ and 300 nm of $Ga_{0.93}In_{0.07}As_{0.98}N_{0.02}$ (laser absorber). The reactor was heated to 650 °C to deposit 200 nm of GaAs (spacer) and 1 µm of $Ga_{0.51}In_{0.49}P$ (upper stop-etch). The reactor was again cooled to 570 °C to grow 50 nm of $Ga_{0.51}In_{0.49}P$ and 100 nm of $Ga_{0.97}In_{0.03}As_{0.99}N_{0.01}$:Se (front contact). The reactor was again heated to 650 °C to grow 200 nm of GaAs:Se (spacers), a bilayer window of 14 nm of $Al_{0.52}In_{0.48}P$:Se and 6 nm of $Ga_{0.51}In_{0.49}P$:Si, 1.8 µm of GaAs:Si (absorber) and 300 nm $Ga_{0.51}In_{0.49}P$:Si:Zn (back surface field). The reactor was cooled to 620 °C to grow 50 nm of $Al_{0.3}Ga_{0.7}As$:C (rear contact), cooled to 250 °C under AsH3, cooled to room temperature under $H_2$, and the sample removed under $N_2$. Specimens were then shipped from NREL to Stanford, contained but unsealed.

**Substrate Polishing**

For crystal ejection specimens, post-growth, the thin film surface was coated with Krylon Workable Fixatif, allowed to dry, and mounted to a handling plate using a heat-softening wax. The rough surface was lapped against a glass plate coated with a 100 nm colloidal $Al_2O_3$ suspension (pH 4, specific gravity 1.14, 20% solids content), then a 70 nm colloidal $SiO_2$ (pH 7.3, specific gravity 1.2, 30% solids content), with materials and data from Eminess Technologies. The specimen was removed from the handler with acetone and blown dry with N2. Final specimen thickness was nominally 250 µm.

Rear Contact Evaporation and Ejection Preparation

Working in a Class 100 cleanroom, after polishing, we cleaved a rectangle off the short edge of the specimen using a diamond scribe and light pressure. The cleaved specimen was rinsed with acetone, isopropanol, deionized (DI) $H_2O$, and dried. The specimen was held in 9:1 (V:V) DI $H_2O$ : 28%-$NH_4OH$ for 30 seconds to remove surface oxides, rinsed in DI $H_2O$, and then dried. Working expeditiously, after drying, the specimen was set onto a glass microscope slide (that had been rinsed in acetone and DI $H_2O$ and dried), and taped to the slide at the specimen corners, with the tape edges oriented 45° to the specimen edges to prevent fracture. The sample was loaded into an e-beam evaporation chamber which was immediately evacuated.

At a base pressure of $7 \times 10^{-7}$ Torr, nominally 500 nm of Au were e-beam evaporated onto the rear contact. Initial evaporation rates were set (by a quartz crystal monitor) to about 5 A $s^{-1}$, the shutter opened, and then the evaporation rate was ramped to between 10 A $s^{-1}$ to 20 A $s^{-1}$ to finish. This was followed immediately by depositing a 40 nm Ti adhesion layer, by again stabilizing at about 5 A s$^{-1}$, opening the shutter, and ramping to 10 A s$^{-1}$ to 15 A s$^{-1}$. The specimen was removed from the glass slide.

We then bonded glass carriers to the Ti surface. The carriers were 4 mm x 4 mm squares diced from glass coverslips (Gold Seal® #1.5, 160 µm to 190 µm thick), and spaced at least 1 mm from a {110} edge or any other carrier. Specimens that were too close to the edges could fail to eject due to diffractive edge effects, and specimens that were too close to each other could fail to eject due to diffractive effects from the Al foil masks used to shield adjacent specimens during ejection. Specimens were always set on cleanroom wipes with the surface of interest facing up to avoid contamination and tearing.

The carriers were rinsed in acetone and dried. A pipette tip was dipped into a puddle UV-curing adhesive (Norland Optical Adhesive 73) and dabbed onto the Ti surface, then the carrier was set onto the adhesive. Capillary forces would draw the adhesive into the space between the carrier and Ti. Bubbles in the adhesive were allowed to burst, or burst manually, before placing the carriers. The specimen was then placed into a MicroWriter ML®3 385 nm UV direct-write lithography system. The coordinates of the four carrier corners were used to define the center of the carrier, and then the adhesive under the carrier was cured with a 4 J cm$^{-2}$ dose. After curing, the carriers were rinsed in acetone and dried to remove wet fillets and any smeared adhesive. The specimen was then placed, carrier side up, under a metal halide UV lamp (LOCTITE® 7411-S UV Flood System) for 3 min to ensure complete curing.

The next step was protecting the optical surface before metal etching. The specimen was set carrier-side down to expose the polished substrate. A 3 mL polyethylene pipette was dipped into a small pool of Shipley S1813 photoresist. The drop that formed on the pipette was dragged along the polished surface, and the pipette could be squeezed to expand a bubble to fill in corners and right up to the edge, to completely coat the surface without photoresist wetting around the edge and onto the metal surface. The photoresist was cured on a Teflon-coated hotplate at around 100 °C. The specimen was removed and placed on a cleanroom wipe to prevent the photoresist from cracking (due to too rapid of a cool-down). The surface was inspected with an optical microscope to ensure there were no cracks or uncovered surfaces.

To remove the metal from around the carriers, a few drops of 49% HF were added to 25 mL of H$_2$O. A flat-bottomed dish with high sidewalls was filled with about 5 mm of a standard I/KI Au etchant. This liquid depth prevented the dark solution from blocking visibility to the bottom of the dish. The specimen was dipped into the HF and gently oscillated. The oxidized Ti surface was etched slowly by the dilute HF, but after an order 10 s delay, the Ti etched in seconds, generating bubbles. The specimen was rinsed DI $H_2O$ and dried with $N_2$. The specimen was then placed, metal-side-up, into the Au etchant and pressed to submerge. The dish was oscillated to swirl the liquid over the surface and prevent precipitant redeposition. When the Au appeared to be gone, the specimen was rinsed in DI $H_2O$ and dried with $N_2$. Finally, the photoresist was rinsed off with acetone, and the specimen was blown dry with $N_2$, and set onto a soft, covered surface for transport out of the cleanroom.

**Pulsed Laser Crystal Ejection**

Crystal ejection was performed with a nominally 2.5 J per pulse, 8 ns to 10 ns FWHM, 1064 nm, Q-switched Nd:YAG laser (Spectra-Physics Quanta-Ray Pro-350). The laser consisted of two oscillating stages and two amplifying stages. The laser was operated at full power in single-shot mode. Typical pulses at the outlet were measured around 1.4 J per pulse via an attenuated thermopile detector and power meter (Maestro, Gentec-EO). The laser pulse was steered with various high damage threshold dielectric mirrors, through a vacuum pinhole spatial filter, to a dichroic mirror that redirected the pulse straight down onto a specimen and a Si CMOS beam imaging array (if installed; Beamage 3.0, Gentec-EO). More system details are given in S3.

After normal startup of the laser system, the flashlamps for all stages were allowed to flash at full power and 10 Hz to 20 Hz to warm up (without the laser firing) while the system was prepared for ejection. All optics were dusted with a manual air blower, and visually inspected for defects and dust. Detector cards, neutral density (ND) filters, and beam imaging were used to check system alignment. The ND filters were kept slightly loose in their housing to prevent temperature increases from breaking the filters. We adjusted the pinhole position and system alignment in an iterative process to maximize power transmission and beam profile symmetry. Typical spatial filter inlet and outlet energies were 1.4 J and 1.2 J, respectively. Then, a specimen was installed above the beam imager on a 2-axis translating stage, by sliding it into a mounted Al foil pocket which acted as both a holder and a mask (to shield nearby specimens from laser energy, as necessary). The substrate side was facing up and the stage could translate parallel to the table. Additional foil strips were used to shield adjacent specimens from the laser pulse, as required.

After specimen mounting, the beam was attenuated by about $10^6$ with ND filters. Pulses were fired and the transmitted images inspected to check alignment of the opaque metal contact, until the specimen was approximately centered on the laser pulse. The beam imager was removed from the optical path. An Al foil-lined, glass beaker was placed below the specimen, and physical, adhesive barriers were installed around and above the specimen, e.g. cardboard and double-sided tape. A facemask was dawned (while wearing a laboratory coat and gloves), and the ND filters were removed. A pulse was fired to eject the specimen. If the specimen did not eject, the specimen was translated (or the foil masks shifted) before firing another pulse, and this process repeated until the specimen ejected or the surface appeared too fouled by surface damage to continue. After finishing the batch, specimens were transported back to the cleanroom.

**Device Finishing**

Ejected specimens were rinsed in acetone and DI $H_2O$ and dried with $N_2$. We cleaned a 22 mm x 22 mm cover glass (Gold Seal® #1.5, 160 µm to 190 µm thick) with acetone and DI $H_2O$ and dried with $N_2$. Each device carrier was adhered near the corner of a cover glass with a dab of UV-curing adhesive (Norland Optical Adhesive 73). The cover glass was then inverted and cured under a metal halide UV lamp (LOCTITE® 7411-S UV Flood System) for about 5 minutes, being careful to prevent the wet adhesive from contacting another surface.

We prepared 2 room temperature solutions for wet etching: $H_3PO_4 : H_2O_2 : H_2O$ (3:4:1 by volume) and full strength HCl. Specimens were gently agitated during all wet etching steps, and it was usually clear when any etch step completed based on expected etch times and visual indications, such as the end of interference-related color changes or the revealing of the rear Au surface. Using tweezers, the specimens were immersed into the (stirred) phosphoric acid solution to dissolve the melt debris. The melt debris dissolved, but was at least a six-component (senory) alloy, as well as polycrystalline and likely oxidized (because ejection was performed in air), so it did not etch like typical single-crystalline, III-V thin films. The melt debris would dissolve, but agitation during etching could also shake off flakes of melt debris which then dissolved into solution. After the GaInP surface was revealed, the specimens were immersed in DI $H_2O$, agitated, and thoroughly dried with $N_2$. The specimen was then immersed in HCl to etch away the GaInP (usually bubbling). When the GaInP was gone, the specimen was immersed in DI $H_2O$ and dried with $N_2$.

Having exposed the device's GaInAsN front contact layer, next we electroplated the front Au contact grid. We minimized edge beading by adhering layers of Kapton® tape on each other until the thickness was around that of the carriers, then cut out mm-scale sections with a razor and bonded them along the outer edge of carriers. This provided a surface onto which the photoresist could flow during spin coating, instead of beading at the edge. The contact surface was heavily doped to form an Ohmic contact with the device grid, which meant that a metal in contact with this film could serve as an electrical lead for electroplating. Such a lead also had to submerge in electroplating solution without contacting the solution, otherwise metal would electroplate onto the metal contact. We used Cu foil cut into mm-wide, cm-long strips. The Cu end needed to push onto the film to avoid getting undercut by photoresist during spin coating, so a razor was pulled along the bottom of the exposed Cu to curl the end of the strip. The Cu strip was adhered to a strip of Kapton® tape such that the curled Cu edge extended slightly past the Kapton®. Another strip of Kapton® tape was adhered to the cover glass, up to the corner of the device. Then the Cu strip was adhered on top of that such that the Cu contacted the corner of the thin films. Tweezers were run along the Cu/Kapton® edges, to help electrically isolate the length of the Cu, and to ensure the Cu stayed in contact with the film during spin coating and electroplating.

We then mounted the specimen to a spin coater vacuum chuck and covered the entire handler with SPR 220-7 (thick, viscous) photoresist, then used the dispensing pipette to gently draw and expel photoresist under the Cu contact at the device corner to flush out any bubbles. Specimens were spun at 3500 revolutions per minute for 45 seconds, then baked on a hotplate for 90 s at 95 °C, then 90 s at 115 °C. Specimens were moved to a cleanroom wipe and allowed to cool. The end of the Cu opposite of the device was cleared of photoresist with a cleanroom wipe and acetone in order to allow metal tweezers to hold the handler while contacting the Cu during electroplating. Devices were set aside in a dark container and allowed to sit for at least 1 hr to allow moisture to absorb into the photoresist before exposure. Contact patterns used 8 µm-wide grids with 250 µm spacing and a 500 µm x 500 µm contact pad at a corner. The pattern was written at 750 mJ $cm^{-2}$ and then developed in MF-26A developer until the resist was observed to clear, then rinsed with DI $H_2O$ and dried with $N_2$.

Contact electroplating was done with a cyanide-free, neutral-pH, Au electroplating solution (Transene TSG-250) and Watts Ni electroplating solution (Transene). Au electroplating used a Pt-coated Ti mesh anode, and Ni electroplating used an unbagged Ni anode. All current densities were calculated using the area of the digital direct-write grid pattern. Solutions were brought to 60 °C on hotplates under mild stirring. The exposed Cu on the edge of our carrier was clamped with normally-closed metal tweezers connected to a current-limited source meter (Keithley). The specimen was immersed in A 9:1 DI $H_2O$:28% $NH_4OH$ solution for 30 s to remove surface oxides, then immersed in DI $H_2O$. The power supply was turned on and the specimen moved into the Watts Ni. Ni was electroplated at 2.5 mA $cm^{-2}$ for 30 seconds. The specimen was immersed in warm DI $H_2O$, then immersed in the Au electroplating solution. Au was electroplated at 2.5 mA $cm^{-2}$ for 2 min then 5 mA $cm^{-2}$ for 5 min. The specimen was removed from the solution, immersed in DI $H_2O$ and dried with $N_2$. The photoresist was removed by rinsing in acetone, DI $H_2O$, and drying with $N_2$. The Cu lead and all Kapton tape were gently peeled off.

To etch away the GaInAsN contact layer (between grid fingers) and expose the AlInP window layer, the specimen was immersed in DI $H_2O$ to wet the surface, then immersed for about 10 seconds in a $NH_4OH : H_2O_2 : H_2O$ (2 :1: 10) solution, then immersed in DI $H_2O$ immediately when the layer colors stopped changing. This minimized undercutting of the Au contacts. Finally, the mesa was defined with photolithography and a series of etches. We again built a Kapton® tape bridge to prepare the specimen for spin coating. Devices were spin coated with Shipley 1818 at 4000 revolutions per minute, then cured on a hotplate for 5 min at 100°C. The mesa area was defined with the ML®3. We heavily exposed everything around a square area, oriented over the contacts, using ~1 J $cm^{-2}$ fluences to minimize photoresist remaining near the edges, and the photoresist was developed in MF-26A and rinsed with DI $H_2O$.

We prepared three solutions, 1:4:3 DI $H_2O$:$H_2O_2$:$H_3PO_4$, concentrated HCl, and DI $H_2O$. For each step, after ensuring that the tweezers and specimen were thoroughly dry, the specimen was immersed in an etchant to remove layers, then rinsed with DI $H_2O$ and dried: AlInP and GaInP window layers in HCl, GaAs absorber in $H_2O$:$H_2O_2$:$H_3PO_4$, GaInP back surface field in HCl, and AlGaAs rear contact in $H_2O$:$H_2O_2$:$H_3PO_4$. The photoresist was removed by rinsing in acetone, isopropanol, DI $H_2O$, and drying with $N_2$, completing the device.

**Solar Cell Testing**

The quantum efficiency was measured on a custom-built instrument based on a tungsten-halogen lamp chopped at 313 Hz, and a 270m monochromator. Data were acquired every 5 nm. The photocurrent at each wavelength was measured with a Stanford Research Systems SR570 low-noise current preamplifier and an SR830 dual-channel lock-in amplifier.

Current-voltage measurements were taken on a custom-built solar simulator based on a xenon lamp, tuned to simulate the AM1.5 G173 direct spectrum at 1000 W/m$^2$ and 25°C. The illumination spectrum was measured with a Spectral Evolution high-speed spectrophoto-radiometer. The intensity was set using a calibrated GaAs reference cell, with a spectral mismatch correction of 0.989. IV curves were measured from forward-to-reverse bias, dwelling 10 ms at each point, and no pre-conditioning was required for these III-V devices. We estimate the relative uncertainty on the efficiency to be ~3%, dominated by the uncertainty in the current density due to fluctuation and non-uniformity in the lamp intensity. Note that the cell performance reported here was not independently confirmed by a certified laboratory, as the cells do not purport to demonstrate record or near-record efficiencies.

Laser-ejected cells were not masked during J-V measurements. Reported cell areas and contact areas for these cells were measured using image analysis of the bright-field optical micrographs (Fiji/ImageJ). The front contact pad was excluded from the measured cell area but the front grid lines were not. All III-V material that was not clearly separated from the main cell area or semi-transparent was included in the total area. For example, in the bright field image in Figure 3, the protruding area on the bottom of the left edge was included, but the red/green etched area just below that protrusion was not. For each cell, repeated image analysis yielded the same area to 0.001 cm$^2$, so we report a relative uncertainty of 1%. For each of the two laser ejected cells, the average of three optical analyses of the total front Au contact area was 8.8e-3 cm$^2$ and 7.4e-3 cm$^2$, respectively. Each reported short circuit current density is the average of the five data points from -0.01 V to 0.01 V.

**Transmission Electron Microscopy**

The TEM specimen was cut from below the front Au contact pad of a finished cell using an FEI Helios Nanolab 600i DualBeam FIB/SEM. The cross-section was welded to a Cu lift-out grid and lifted out using an Omniprobe AutoProbe 200 in situ specimen lift-out system. Bright field imaging was performed using an FEI Titan 80-300 transmission electron microscope operated at a 300kV accelerating voltage. The aberration corrector in the objective lens was tuned before imaging.

**X-Ray Diffraction**

A section of the as-grown specimen was selectively etched to reveal the GaAs|GaInAsN|GaInP layers grown beneath device layers. X-ray diffraction was performed with a PANalytical X'Pert Pro X-ray Diffractometer using Cu K$_{\alpha 1}$ radiation, $\lambda$=1.5406 Å. We performed reciprocal space mapping on the (004) and (-2-24) Bragg peaks (using the grazing emergence geometry to improve resolution). All scans used a hybrid monochrometer with a 5 mm mask and a 1/4˚ slit for the incident optics, and an unattenuated PIXcel 1D detector with a Soller slit for the diffracted optics. More X-ray diffraction details are presented in S2.

**Photoluminescence**

Room temperature photoluminescence was performed on the X-Ray Diffraction specimen, using excitation wavelength $\lambda$=532 nm and power P=6.3 mW (Ophir). A microscope (Witec) focused the exciting light with a 50x, 10.1 mm working distance, numerical aperture NA=0.55, infinity corrected objective (Nikon LU Plan). The objective collected the photoluminescence, which passed through a 532 nm laser filter and any installed long pass filter before fiber coupling into a multimode fiber (NA=0.12), connected to a second multimode fiber (NA=0.39), into a spectrometer with 500 nm blazed grating (ACTON SpectraPro 2300i). Spectra were collected using a liquid N$_2$ cooled, unstrained InGaAs detector (Princeton Instruments), operating at temperature T = -100 ˚C, with a detection range of 800-1700 nm. The spectrometer was set to collect the range of interest, background checked, then Ne light was coupled into the detector to collect peak positions for calibration check, followed by specimen spectra collection. More photoluminescence details are presented in S2.

# Supplementary Information: Pulsed Laser Ejection of Single-Crystalline III-V Solar Cells From GaAs Substrates

Benjamin A. Reeves,[1] Myles A. Steiner,[2] Thomas E. Carver,[3] Ze Zhang,[4]
Aaron M. Lindenberg,[1,5] Bruce M. Clemens[1*]

1. Department of Materials Science and Engineering, Stanford University, Stanford, CA.

2. National Renewable Energy Laboratory, Golden, CO.

3. Stanford Nano Shared Facilities, Stanford University, Stanford, CA.

4. Department of Mechanical Engineering, Stanford University, Stanford, CA.

5. Stanford Institute for Materials and Energy Sciences, SLAC National Accelerator Laboratory, Menlo Park, CA 94305

* To whom correspondence should be addressed: bmc@stanford.edu

## S1: *In Situ* Curvature and Reflectivity

Figure S1 shows the *in situ* wafer curvature and reflectance measurements from a k-space multibeam optical stress sensor system, as described in SR1 and SR2. The red and blue curves in the top panel are approximately aligned with the [110] and [-110] directions. The straight and nearly flat lines indicate good lattice-matching in all layers. The vertical gray lines delineate the individual layers; some layers are indicated with text. The black curve in the bottom panel shows the reflectance fringes during growth, indicating a specular growth surface.

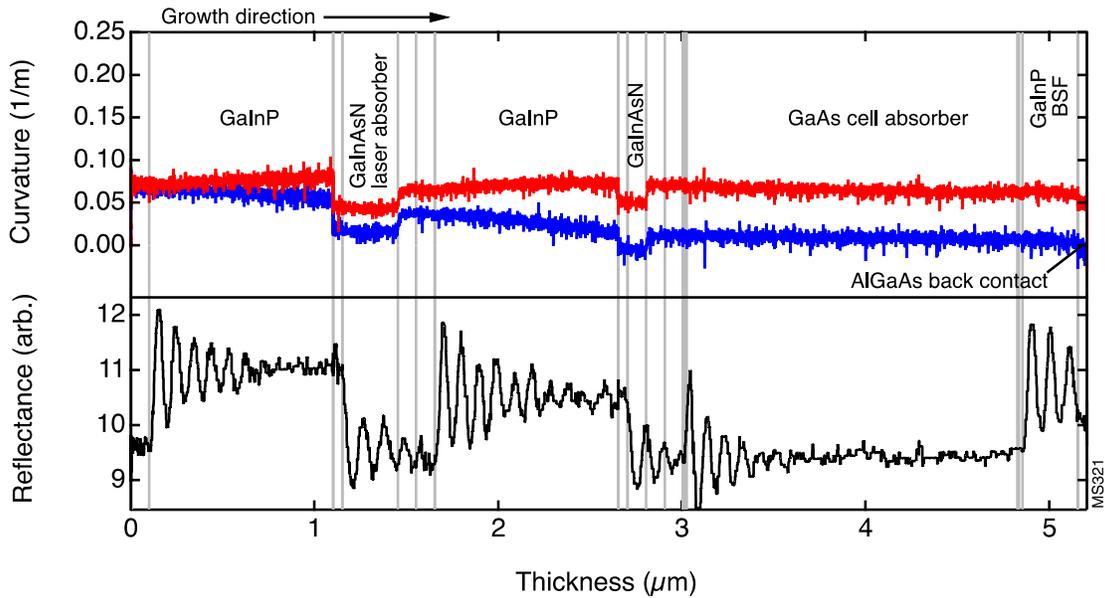

**Figure S1:** The *in situ* MOVPE wafer reflectance and curvature measurements for the laser ejected solar cell synthesis, as a function of nominal layer thickness. Layers are delineated by vertical lines—including two 100 nm GaAs layers between the GaInAsN laser absorber and GaInP, grown with different Ga precursors—and some are labelled.

## S2: GaInAsN Absorber Characterization

We estimated the in-plane lattice parameters and band gap of the GaInAsN absorber layer using X-ray diffraction and room temperature photoluminescence. We find that the nominally $Ga_{0.93}In_{0.07}As_{0.98}N_{0.02}$ quaternary absorber is lattice matched to GaAs to within order 0.001 Å, with a band gap of approximately 1.07 eV.

The as-grown specimen was etched in room temperature $H_3PO_4:H_2O_2:H_2O$ (3:4:1 by volume) and HCl, to etch III-As and III-P layers, respectively. After each step, the specimen was moved into DI H2O, and the specimen and tweezers were dried with dry $N_2$ before immersing in HCl. The AlGaAs rear contact, GaInP BSF, GaAs absorber, GaInP-AlInP windows, GaInAsN front contact, and GaInP etch stop were etched away to reveal the 200 nm GaAs above the 300 nm GaInAsN absorber.

X-ray diffraction was performed with a PANalytical X'Pert Pro X-ray Diffractometer using Cu $K_{\alpha 1}$ radiation, $\lambda=1.5406$ Å. After zeroing $2\theta$ and aligning $z$ and $\omega$ to place the sample into and parallel to the beam, the (004) Bragg peak was found tilted 2.1° from the expected $\omega$, consistent with the 2° (111)B miscut specified by the vendor. We performed reciprocal space mapping on the (004) Bragg peak by scanning in $2\theta$ and $\omega$. Post-collection, we shifted all $\omega$ values for the (004) map by -.023° due to forgetting the numerical offset during fine alignment. Then, we aligned, calibrated to, and mapped the (-2-24) Bragg peak, using the grazing emergence geometry to improve resolution. All scans used a hybrid monochromer with a 5 mm mask and a 1/4° slit for the incident optics, and an unattenuated PIXcel 1D detector with a Soller slit for the diffracted optics. The reciprocal space maps are shown in Figure S2. The vertical alignment of the (004) and (-2-24) peaks indicates that the GaInAsN absorber is pseudomorphic on the GaAs substrate. Although the GaInP, GaAs, and GaInAsN peaks are not distinguished, the collection of peaks within the scattering vectors $|q|$ = 4.440 Å$^{-1}$ and 4.445 Å$^{-1}$, via the Bragg condition (at $l$=4)

$$a_\perp = \frac{8\pi}{|q|}$$

implies that the (strained) out-of-plane lattice parameters $a_\perp$ are equal to within order 0.001 Å.

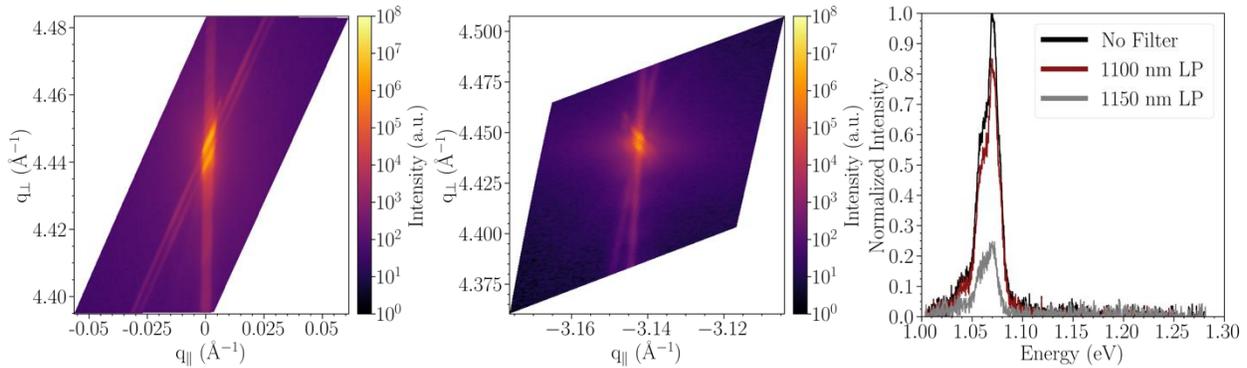

**Figure S2:** The composition of the absorber layer was estimated via X-ray diffraction reciprocal space maps (RSM) and room temperature photoluminescence, after etching away the device layers and upper etch stop from a solar cell specimen. The measured specimen, from the direction of the X-rays, was nominally 200 nm GaAs | 300 nm GaInAsN | 1 μm GaInP | GaAs substrate. Left, center) RSM of (004) and (-2-24) (grazing emergence) Bragg peaks. The peaks are vertically aligned to within the resolution of the experiment, indicating pseudomorphic GaInP and GaInAsN layers. Right) Room temperature photoluminescence spectra with either an 1100 nm or 1150 nm longpass filter inserted between the specimen and spectrometer. We report the band gap as the peak of the unfiltered signal, 1.07 eV ± 0.01 eV, where the uncertainty is the approximate FWHM of measured Ne spectral lines.

Room temperature photoluminescence was performed on the etched specimen. The excitation laser used wavelength λ=532 nm and power P=6.3 mW (as measured by an Ophir power meter at the specimen plane). Using a microscope (Witec), the exciting light was focused with a 50x, 10.1 mm working distance, numerical aperture NA=0.55, infinity corrected objective (Nikon LU Plan), with its optical axis normal to the thin film surface. The objective also collected the photoluminescence, which passed through a 532 nm laser filter before entering a fiber. We connected this fiber (multimode, NA=0.12) to a second fiber (multimode, NA=0.39), which was mounted to a ACTON SpectraPro 2300i spectrometer. The spectrometer utilized a 500 nm blazed grating, and a liquid $N_2$ cooled, unstrained InGaAs detector (Princeton Instruments), operating at temperature T = -100 ˚C, with a detection range of 800-1700 nm.

After setting the spectrometer to collect the spectral range of interest, light from a Ne lamp was directed through the second fiber into the spectrometer, to check the instrument calibration based on NIST reported lines of Ne. Based on these peaks, the GaInAsN spectra were red shifted 3 nm post-collection, and we report the ~0.01 eV FWHM of the calibration peaks as the measurement uncertainty. The fiber was then reconnected to the photoluminescence system, and the GaInAsN spectra collected (Fig. S2). The signal responded as expected when filtering the photoluminescence signal by either an 1100 nm (1.13 eV) or 1150 nm (1.08 eV) longpass filter. Two additional sampled points—collected from the same ~ 2 mm wide sliver that was cleaved from the as-received MOVPE specimen—gave virtually identical spectra (not shown). We report the band gap as the peak of the photoluminescence signal, 1.07 ± 0.01 eV. Based on lattice-matching, the ~1.07 eV band gap, and the work in Ref. SR3, we estimate 2% N content.

Prior work (SR4) ejected GaAs thin films using lattice-matched GaInAsN with a band gap around 0.8 eV (~4% N). Although our measurements did not capture the band gap variability across the MOVPE specimen, we hypothesize that crystal ejection is generally possible with photon energies very close to the direct band gap energy of the absorber (although lower absorption coefficients may significantly increase critical ejection fluence, and lead to unwanted substrate or device damage). This hypothesis is based on the sharp increase of the absorption coefficient with increasing photon energy near the band edge of direct band gap semiconductors (SR5): The absorption coefficient typically rises to nearly 1e4 $cm^{-1}$ at the band edge and remains within a few $10^4$ $cm^{-1}$ for several 0.1 eV above the band gap.

## S3: Laser System Diagram

Figure S3 shows a schematic of the laser system used in this work.

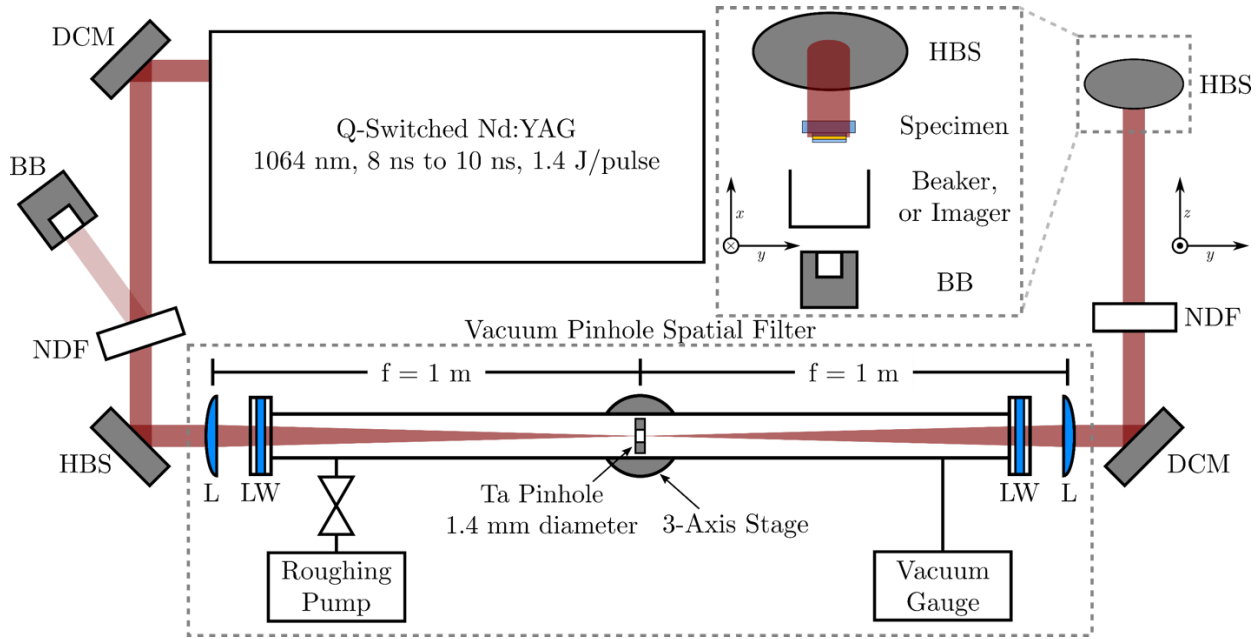

**Figure S3**: A functional schematic of the laser system used in this work. The plano-convex lens at the exit of the spatial filter was approximately 1 m from the pinhole, but located after the respective dichroic mirror, due to optical table space.

A Q-switched Nd:YAG laser (Spectra-Physics Quanta-Ray PRO-350) generated an 8-10 ns full width half maximum, 1064 nm laser pulse. All optical elements, except for the absorbing neutral density filters, had anti-reflection coatings for our laser wavelength and sufficiently high damage thresholds. Beam blocks (BB) were used at any significant source of transmission or reflection. The laser pulse was steered with a dichroic mirror (DCM), through any installed neutral density filters (NDF), and a 1064 nm/532 nm harmonic beamsplitter (HBS) steered the laser pulse onto the optical axis of a vacuum pinhole spatial filter. The harmonic beamsplitter was used to reflect the 1064 nm pulse, but transmit visible light, so that ablation around the metal pinhole could be imaged through the harmonic beamsplitter during major system alignments.

From the beamsplitter, a lens (L) with nominal focal length f = 1000 mm focused the laser into a vacuum spatial filter. The spatial filter consisted of a ~2 m long stainless steel tube under roughing vacuum, with pressure of approximately 10 mtorr as read from an installed digital pressure gauge. The ends were sealed with laser windows (LW; Thorlabs). The 1.4 mm pinhole was drilled through the thin dimension of a ~1 cm x 3 cm x 1 mm Ta sheet, and mounted within the chamber with the pinhole normal to the optical axis, approximately centered lengthwise and within the circular cross section. The spatial filter assembly was bolted to the optical table via a 3-axis translating stage, with approximately 1" of travel along each axis, and with the pinhole located approximately at the nominal focal length.

After exiting the pinhole, the filtered pulse was steered by a dichroic mirror to a collimating lens with f = 1 m. The beam path between the lens and the pinhole was approximately 1 m, but lens was placed after the steering mirror to fit within available table space. The beam passed through additional neutral density filters and onto a power meter (if installed), and was steered down into the plane of the optical table with another harmonic beamsplitter. A beam imager (BEAMAGE-3.0, Gentec-EO) could be inserted normal to the beam downstream of the specimen plane. Neutral density filters were used for beam characterization and specimen alignment, but were all removed during crystal ejection.

Within the laser, the laser pulses are generated with two oscillating stages, and then amplified by two amplifying stages. To avoid pulse gating or subjecting system components to 10-20 Hz, full-energy pulses, the laser system was operated in single-shot mode. However, the power and divergence of the pulse depended on the thermal history of the amplifying stages. We therefore waited at least 30 seconds between each shot to reduce shot-to-shot variability.

The imager used to capture the characteristic beam profile shown in Fig. 1h did not capture all of the pulse, and hence could not be used with the total measured pulse energy $E$ to create a fluence map. To normalize this image to the unknown captured energy $E'$, we analyzed the captured detector image to estimated the total detector counts $C$ that would exist in an entire beam image. This was done by integrating imaged counts $C'$ within a sector spanning an angle $\theta$ and radius $r$, whose radius went from the pulse center to background levels everywhere in $\theta$. Then, by assuming that the sector was representative of the entire radial profile, we used $C=(2\pi/\theta)C'$ to estimate total counts $C$. The ratio of $C$ to the sum of all imaged counts for three estimates was 1.17 to 1.24, so we used $E'=1.0$ J to map fluence instead of $E=1.2$ J.

**S4: Substrate images showing single/multi-pulse ejection, and dark field etch results**

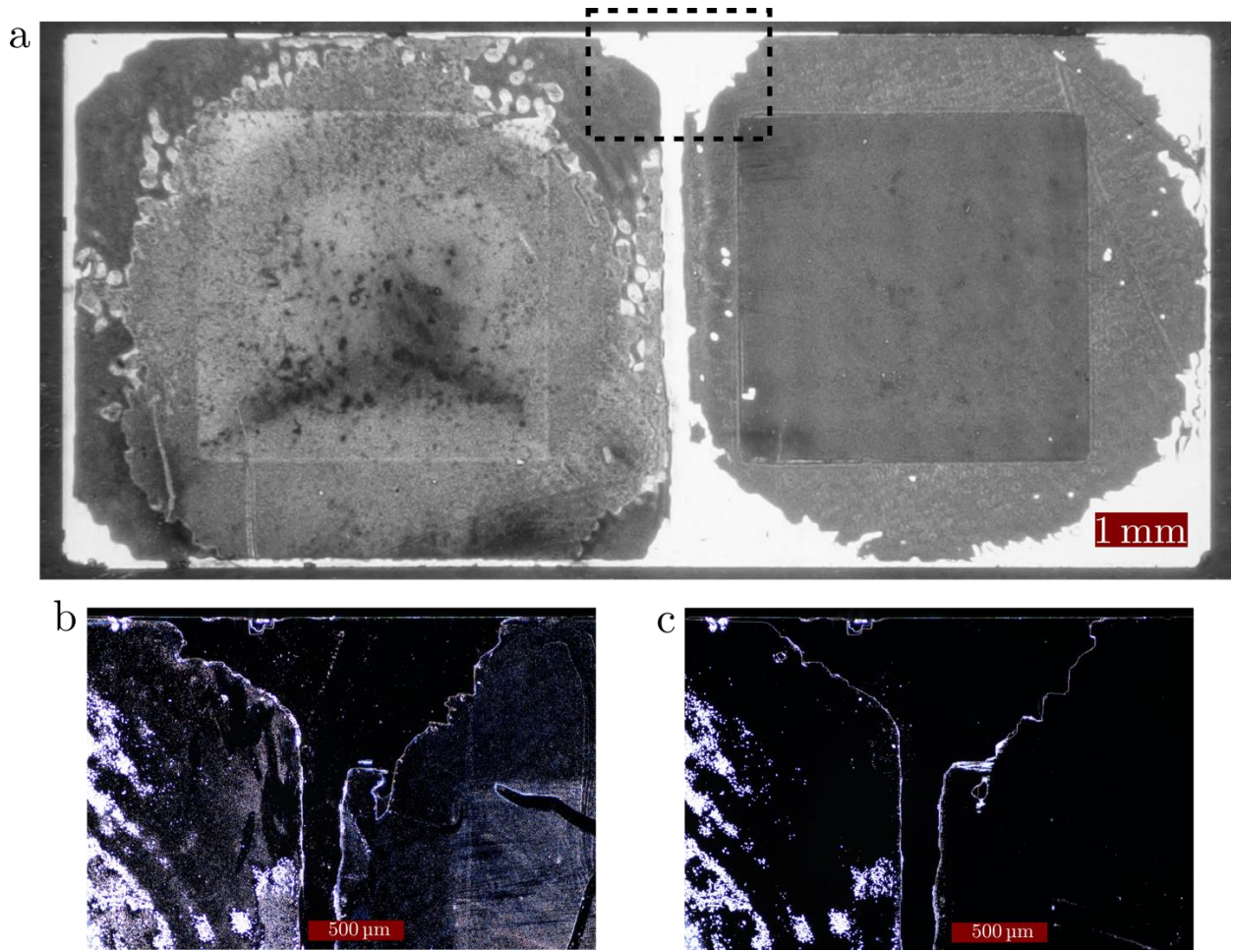

**Figure S4:** a) A grayscale, bright field optical micrograph, showing the thin film side of a specimen piece after pulsed laser ejection of the bonded solar cell layers. The image shows the result of multi-shot and single-shot device ejection on the left and right, respectively, before substrate etching. White specks on the right are either unejected film, or debris that landed on the surface after ejection. b-c) Dark field optical micrographs around the area indicated by the dashed rectangle in a, before and after surface treatments, respectively.

Figure S4a is a gray-scale, bright field optical micrograph of an as-grown film after device ejection and no wet etching. The device on the left was ejected with multiple laser pulses, due to system variability or small misalignments. The device on the right was ejected with a single laser pulse. We then cleaned the substrate using a $H_3PO_4:H_2O_2:H_2O$ (3:4:1) etch, a gentle wipe with a water-detergent cotton swab, and another etch in HCl. Figures S4b and S4c show dark-field optical micrographs before and after this treatment, respectively, of the top-middle portion of the specimen as indicated in Figure S4a. We find a surprisingly clean surface on the single-shot specimen, even though experiments to date have been in air, and with high spatial-variance in pulse fluence. These results are promising preliminary findings for future optimization work, namely with varied etch stop thicknesses, pulse durations, and fluence spatial gradients.

## S5: Additional optical micrographs

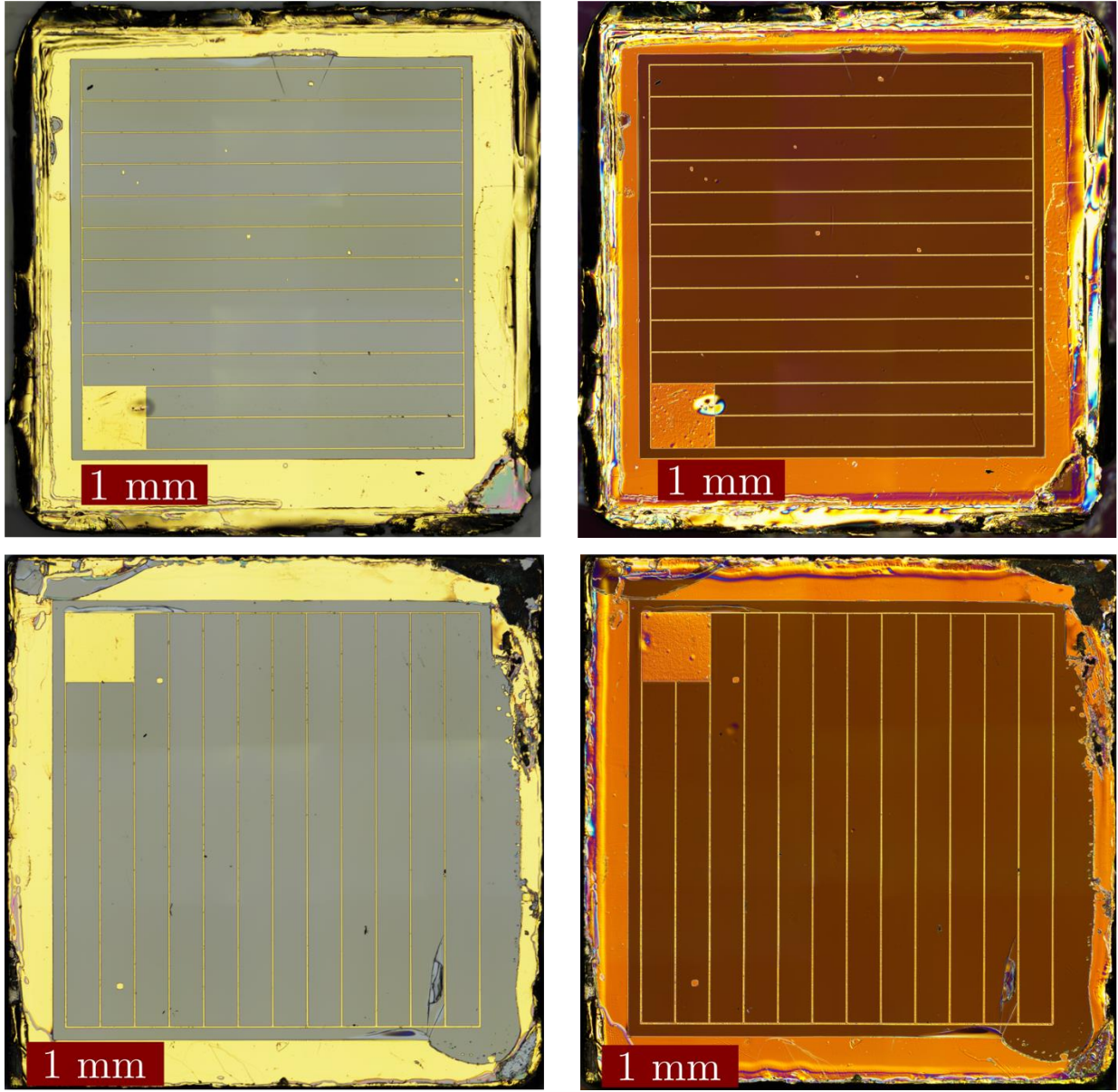

**Figure S5:** Bright field and differential interference contrast micrographs of two additional, ejected specimens. The top figures show the second successful solar cell described in Figure 2; the two cracks on top of the cell were caused by photoresist curing after ejection. The bottom figures show an additional cell that was shunted during testing; the crack on the bottom left was caused by handling after fabrication, but before JV testing. The vertical and horizontal discolored areas resulted from image stitching.

**Supplementary Information References**